\documentclass[aps,prl,amsmath,amssymb,twocolumn,floatfix,10pt,showpacs]{revtex4-1}
\usepackage{graphicx}
\usepackage{graphics}

\usepackage{amsmath}
\usepackage{amssymb}
\usepackage{amsfonts}
\usepackage{dsfont}
\usepackage{sidecap}
\usepackage{hyperref}
\usepackage{color}

\begin{document}

\title{Electric-Field Induced Majorana Fermions in Armchair Carbon Nanotubes}
\author{Jelena Klinovaja}
\author{Suhas Gangadharaiah}
\author{Daniel Loss}
\affiliation{Department of Physics, University of Basel, Klingelbergstrasse 82, 4056 Basel,
Switzerland}
\date{\today}

\begin{abstract}

We consider theoretically an armchair Carbon nanotube (CNT)
in the presence of an electric field and in contact with an $s$-wave
superconductor. We show that the proximity effect
opens up superconducting gaps in the CNT of different strengths for the
exterior and interior branches of the two Dirac points.
For strong proximity induced superconductivity the interior gap can be of
the $p$-wave type, while the
exterior gap can be tuned by the electric field to be of the $s$-wave type.
Such a setup supports a single Majorana bound state at each end of the
CNT.
In the case of a weak proximity induced
superconductivity, the gaps in both branches are of the $p$-wave type.
However, the temperature
can be chosen in such a way that the smallest gap is effectively closed.
Using renormalization group techniques we show that the
Majorana bound states exist even
after taking into account electron-electron interactions.
\end{abstract}

\maketitle
{\it Introduction.} 
Majorana fermions in solid state systems have attracted considerable attention recently \cite{fu,Akhmerov,lutchyn_majorana_wire_2010, oreg_majorana_wire_2010, potter_majoranas_2011, alicea_majoranas_2010,Qi, suhas_majorana,stoudenmire,lutchyn}.
In particular, the possibility of realizing them as bound states at the ends 
of semiconducting nanowires  in the proximity of an $s$-wave bulk superconductor
has led  to much activity. Such setups require a Zeeman splitting, typically generated by
an external magnetic field \cite{footnote_Overhauser_field},  
that must be larger than the proximity induced gap to induce an effective $p$-wave superconductor
in the topological phase. Such a magnetic field, however, tends to destroy the gap in the bulk superconductor itself,
and thus a delicate balance must be found~\cite{stanescu}. It is therefore very desirable to search for Majorana-scenarios  which do not require magnetic fields.

One  of the prerequisites for a Majorana bound end state (MBS)  is the existence of helical modes, i.e.  modes which carry opposite spins in opposite directions.
It has been shown recently that such helical states are induced in Carbon nanotubes (CNT) via  spin-orbit interaction (SOI)  by an external {\it electric} field \cite{klinovaja_helical_modes_2011,klinovaja_cnt_phys_rev_B}. This mechanism works optimally for a special class of metallic CNTs: armchair CNTs $\rm (N, N)$.
This class is characterized by a spin-degenerate low-energy spectrum around  the two inequivalent Dirac points, $K$ and $K^\prime$.
This degeneracy can be lifted by an electric field which gives then rise to helical modes.

\begin{figure}[tbhp]
\includegraphics[width=0.5 \textwidth]{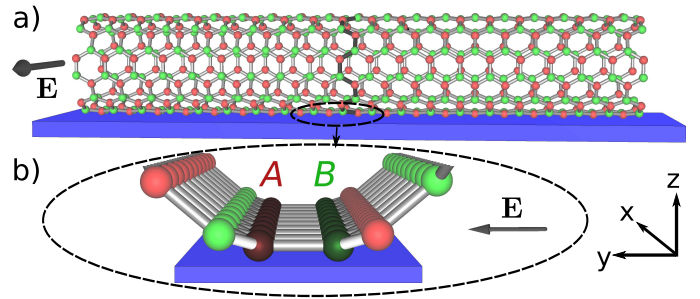}
\caption{(a) An armchair nanotube (cylinder) is placed on top of a superconductor (blue slab). The $x$-axis points along the nanotube. An electric field $E$ is applied perpendicular to the nanotube in the $y$-direction. There are two non-equivalent lattice sites: $A$ (light red) and $B$ (light green). (b)  The distances between the superconductor surface and the atoms of  sublattice $A$ (dark-red row) and of sublattice $B$ (dark-green row)  are assumed to be the same. Thus, the tunneling amplitudes  to the different sublattices are (nearly) equal.}
\label{system}
\end{figure}

\begin{figure}[tbhp]
\includegraphics[width=250pt]{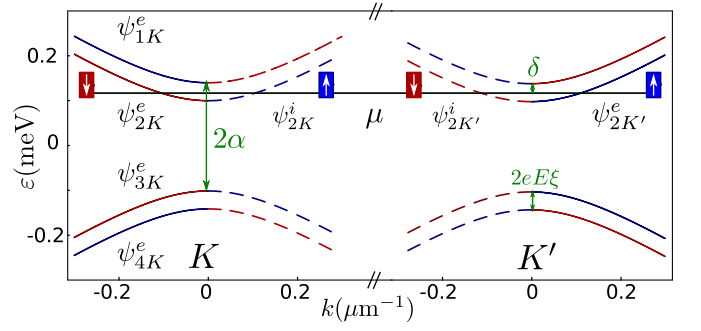}
\caption{The energy spectrum around the Dirac points $K$, $K^\prime$ for a (10,10)-CNT  in an electric field $E=1\ \rm V/\rm nm$, which consists of exterior (full line) and interior (dashed line) branches.
Each branch of the spectrum is characterized by the sign of the spin projection along the $z$-axis  $\left<S_z\right>$ (red: spin down, blue: spin up). The Fermi level $\mu$ lies inside the gap given by $2eE\xi$, and  $\delta=eE\xi+\alpha-\mu$. }
\label{fig_spectrum}
\end{figure}

However, when putting the CNT in contact with an $s$-wave superconductor (see Fig.~\ref{system}) with the goal to generate MBS  the following problem is encountered.
The two Dirac points $K$ and $K^\prime$ are Kramers partners (see Fig.~\ref{fig_spectrum}) and thus the superconducting pairing induced  via the proximity effect will involve both of them, {\it i.e.} left (right)-moving electrons from the branch at $K$ get paired with the right (left)-moving electrons from the   branch at $K^\prime$ to form an $s$-wave Cooper pair with zero total momentum.
 This then immediately implies that there will be two superconducting gaps, an `exterior' one, $\Delta_e$, and an `interior' one, $\Delta_i$.  
 Thus, in general, we expect {\it two} MBS at each end of the CNT (i.e. four in total). This, however, is problematic as the Majorana pair at a given end can easily combine to form a single fermion by some local perturbation.
Thus, the question then
arises whether  it is possible  to find a regime with only one MBS 
at each end~\cite{footnote_k=0}.
As we will show, the answer is affirmative but under
  rather stringent conditions. One of them requires a comparable tunnel
 coupling of the A and B sublattices of the CNT 
 to the superconductor, see Fig.~\ref{system}.
  Using the interference mechanism first described by  Le Hur
 {\it et al.} \cite{le_hur_2008}, we will show
that for this particular case the exterior (interior) gap  $\Delta_e$ ($\Delta_i$) gets enhanced (suppressed) due to
constructive (destructive) interference in the tunneling process.  If $\Delta_{e/i}$, is smaller (larger) than the gap opened by the electric field, then the coupling between the two Dirac points is of $p$-wave type (s-wave type). This   leads to two 
 regimes for MBS. 
 The first one involves a scenario wherein only one of two branches has a $p$-wave gap, thus giving rise to only one MBS 
at each end of the CNT. In a second regime, where both branches have a potential for $p$-wave pairing, 
 the temperature $T$ can be choosen to lie between the two gaps $\Delta_e$ and $\Delta_i$, so that  only the exterior branches will go fully superconducting, whereas the interior branches stay normal. Again, a single pair of MBS in the CNT emerges. 
We further investigate the effect of interactions on the MBS. This is particularly important for the second regime due to the presence of  gapless states from the interior branches
that could be  harmful to the MBS.
However, using bosonization techniques 
we will dispel these concerns and show that 
for screened interactions  the MBS remain stable although they can get substantially delocalized similar to 
the simpler case of  
Rashba wires
\cite{suhas_majorana}.

{\it Nanotube spectrum.}
We consider an armchair CNT in the presence of an electric field $E$ applied perpendicular to the CNT axis (see Fig. \ref{system}). Taking into account the 
spin-orbit  interaction the low-energy sector is described by an effective Hamiltonian around the Dirac points given by \cite{klinovaja_helical_modes_2011}
\begin{equation}
 H= \hbar \upsilon_F k \tau_3 \sigma_2 + \tau_3 e E \xi  S_z \sigma_2+\alpha  S_x\sigma_1,
\label{armchair}
\end{equation}
where $k$ is the momentum along the nanotube axis taken from the Dirac point,
$\sigma_{i}$ is the Pauli matrix on the sublattice space $(A,B)$ associated with the honeycomb unit cell, and $S_{i}$ is the spin operator with eigenvalues $\pm1$. The Pauli matrix $\tau_i$ acts on the $K,K^\prime$-subspace.
Here, $v_F \simeq 10^6$ m/s is the Fermi velocity, and the parameter $\alpha$ arises from the interplay between SOI and curvature effects \cite{izumida_soi_cnt_2009, jeong_soi_cnt_2009, klinovaja_helical_modes_2011}. In the framework of the tight-binding model, $\alpha=-0.08\,{\rm meV}/R[{\rm nm}]$, where $R$ is the radius of the CNT \cite{klinovaja_helical_modes_2011}. The parameter $\xi \simeq 2 \times 10^{-5} {\rm nm}$ is given by a combination of hopping matrix  elements, on-site dipole moment, and SOI \cite{klinovaja_cnt_phys_rev_B}.

The spectrum given by $H$ (Eq. \ref{armchair}) consists of four branches (see Fig. \ref{fig_spectrum})
\begin{equation}
\varepsilon_n (k) = \pm e E \xi \pm \sqrt{\alpha^2 + (\hbar \upsilon_F k )^2}
\label{eq:spectrum_armchair}
\end{equation}
for each Dirac point.
In the following, we label the four branches  by $n=1,...,4$. For each $k$, $n=1$ corresponds to the highest eigenvalue  and $n=4$ to the lowest. The remarkable feature of the spectrum is the existence of helical modes, which carry opposite spins in opposite directions. The average value of the spin in the direction parallel to the axis of the CNT ($\left<S_x\right>$) or parallel to the applied electric field ($\left<S_y\right>$) is equal to zero. The projection of the spin along the $z$-direction is equal to $\left<S_z\right>=\sin  \zeta $, where $\zeta$ is defined by $\zeta=\arcsin( \hbar \upsilon_F  k/ \sqrt{\alpha^2+(\hbar \upsilon_F  k)^2})$ and depends on the wavevectror $k$. Note that the eigenvectors $\psi^{e/i}_{nK}$ and $\psi^{e/i}_{nK^{\prime}}$ are independent of the electric field $E$.
For a (10,10)-CNT in an electric field of strength $1 \rm V/\rm nm$, and with a Fermi level $\mu$ tuned between the two lowest electronic states polarizations close to $90 \%$ can be reached \cite{klinovaja_cnt_phys_rev_B}.

{\it Proximity induced superconductivity.} If a CNT is in contact with an $s$-wave bulk superconductor, then the proximity effect induces superconductivity also in the CNT which at the BCS mean-field level is described by
\begin{equation}
\sum_{i,j,i',j',s}(\Delta_d c^\dagger_{i p_r s} c^\dagger_{j p_r \bar s} + \Delta_n c^\dagger_{i' p_r s} c^\dagger_{j' p_r \bar s})+h.c.,
\label{proximity}
\end{equation}
where we concentrate on the contribution coming from the $\pi$-bands formed by the radial $p_r$-orbitals \cite{Dresselhaus, klinovaja_cnt_phys_rev_B}. Here,
 $c_{i p_r s}^{(\dagger)}$ are the standard fermionic annihilation (creation) operators, with  $s$ and $\bar s$ denoting opposite  spin states \cite{klinovaja_cnt_phys_rev_B}.
The sum runs over atoms which are in contact with the bulk superconductor: $i$ and $j$ belong to the same sublattice,  whereas $i'$ and $j'$ belong to different sublattices.
Generically, one can assume  that the lattice constant of the superconducting material is not commensurate with the one of graphene. The CNT is placed in such a way that the distance from the superconducting surface to the $A$ and $B$ atoms is  the same (see Fig. 1), which is satisfied for the special case of armchair CNTs.  This ensures equal probability amplitude for tunneling to either sublattice. Since the phase of the superconducting order parameter $\Delta_{d/n}$ can be chosen arbitrary, we assume them to be real. The coupling terms in Eq.~(\ref{proximity}) conserve momentum, so they pair Kramers partners from the opposite Dirac cones.
The process in which the Cooper pair tunnels from the superconductor to either one of  the sublattice  $\sigma$  is written as
\begin{equation}
\Delta_d \sum_{\sigma,s,\kappa}\text{sgn}(\sigma \bar{s}) \psi_{\sigma s \kappa}^\dagger \psi_{\sigma \bar{s} \bar{\kappa}}^\dagger +h.c. \, ,
\label{tun_sym}
\end{equation}
where the indices $\kappa$ and $\bar{\kappa}$ denote opposite Dirac points. The operators $\psi_{\sigma s \kappa}$ and $c_{i p_r s}$ are connected  via Fourier transformation \cite{klinovaja_cnt_phys_rev_B}.
The pairing term between electrons in different sublattices are
\begin{equation}
i \Delta_n  \sum_{\sigma,s,\kappa}\text{sgn}(\bar{s})  \psi_{\sigma s \kappa}^\dagger \psi_{\bar{\sigma}\, \bar{s} \bar{\kappa}}^\dagger +h.c.
\label{tun_nonsym}
\end{equation}

To simplify the notation we introduce Pauli matrices $\eta_i$ which act on the particle-hole subspace, and we work in the  basis 
$({\bf\Psi}, {\bf\Psi}^\dagger)$, with ${\bf \Psi}=
( \psi_{A \uparrow K}, \psi_{B \uparrow K},  \psi_{A \downarrow K} , \psi_{B \downarrow K},\psi_{A \uparrow K^{\prime}}, \psi_{B \uparrow K^{\prime}},  \psi_{A \downarrow K^{\prime}}, \psi_{B \downarrow K^{\prime}})$.
This allows us to rewrite Eqs. (\ref{tun_sym}) and (\ref{tun_nonsym}) in a compact form
\begin{equation}
H_{sc}=  - \eta_2 \tau_1 S_z \Delta_d \ \sigma_3 + \eta_1 \tau_1 S_z \Delta_n \ \sigma_1.
\label{ham_superconductor}
\end{equation}
In the same particle-hole basis, the effective Hamiltonian Eq.~(\ref{armchair}) can be rewritten  as
\begin{equation}
H= \frac{1}{2}(\hbar \upsilon_F k \sigma_2 + \gamma \eta_3 \tau_3 S_z \sigma_2 + \alpha  \eta_3 S_x\sigma_1).
\label{ham_arm_with_eta}
\end{equation}
To express the coupling between the different energy states in a canonical
form we work in the basis of eigenvectors $\{\psi^{e}_{nK}, \psi^{i}_{nK}, \psi^{e}_{nK^\prime}, \psi^{i}_{nK^\prime}\}$.
For the states at the Fermi level, $n=2$,  $H_{sc}$ takes the form
\begin{equation}
\sum_{l=e,i}\Delta_{l}  (\psi^{l}_{2K^\prime} \psi^{l}_{2K}-\psi^{l}_{2K}\psi^{l}_{2K^\prime})  +h.c.\, ,
\end{equation}
with different coupling strengths for the exterior ($e$) and interior ($i$) branches,
\begin{equation}
\Delta_{e/i}=\Delta_d \pm \Delta_n \left |\sin \zeta \right |.
\label{super_coupling}
\end{equation}
We note that the sign reflects the constructive and destructive  interference, {\em resp.}, in the tunneling process from the bulk-superconductor into the CNT \cite{le_hur_2008}.
We neglect the term
$\Delta_n  \cos \zeta$ characterizing the coupling
between $\psi^{e/i}_{2K}$ and $\psi^{e/i}_{4K^\prime}$ which are separated by the particle-hole gap $2\alpha$, see Fig.~\ref{fig_spectrum}. In the following we consider the limit of equal diagonal and non-diagonal parameters, {\em i.e.}, $\Delta_d=\Delta_n$. We note that for $k\gg \alpha/ \hbar \upsilon_F$ the coupling between the interior branches is close to zero and that between the exterior branches is equal to $2\Delta_d$. We show that this asymmetry in the coupling strengths is crucial for the existence of Majorana bound states in CNTs.

{\it Majorana bound states.}
Next, we 
obtain the MBS following the derivation of Ref.~\cite{suhas_majorana}.
For illustrative purposes we derive the bound states that arise by considering  the exterior branches.
The field corresponding to the exterior branch is defined as, $\psi_{e}(x) = \psi_{2K}^R(x) e^{i(k_F+K)x} + \psi_{2K'}^L(x) e^{-i(k_F+K)x},$ where $\psi_{2K}^R(x)$ and $\psi_{2K'}^L(x)$ are the slowly moving
right and left components about  the $K$ and $K'$ points, {\em resp.}
\begin{figure}[!hbtp]
\centering
\includegraphics[width=240pt]{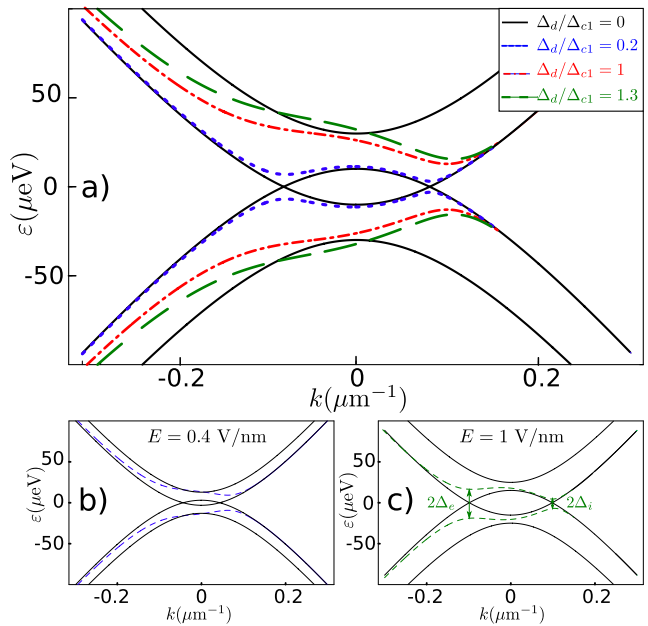}
\caption{The particle-hole spectrum of a CNT (10,10) in the presence of an electric field $E$
with the Fermi level 
 tuned inside the energy gap
between the two upper branches  (solid black lines). All energies are counted from the Fermi level $\mu = 0.11\ \rm meV$ (see Fig.~\ref{fig_spectrum}).
By proximity effect, superconducting gaps $\Delta_{e,i}$ are opened at the Fermi points.
(a) We keep $E$ constant at $1\ \rm V/nm$ and change the strength of the induced superconductivity.  For $\Delta_d=5\ \rm \mu eV<\Delta_{c1}$ both exterior and interior branches are in the $p$-wave phase (dotted blue line). At the critical value $\Delta_d=23 \ \rm \mu eV=\Delta_{c1}$ the gaps induced by the proximity effect and by the electric field are equal (dot-dashed red line). For $\Delta_d=30 \ \rm \mu eV>\Delta_{c1}$ only the interior branch is in the $p$-wave phase (dashed green line).
Keeping  $\Delta_d$ constant at $11\ \mu eV$ and changing $E$, one goes from a regime [dashed blue line in (b)] with $E=0.4\ \rm V/nm<E_{c1}=0.6 \ \rm V/nm$ in which only the interior branch is in the $p$-wave phase to a regime [dashed green line (c)] with $E=1\ \rm V/nm>E_{c1}=0.6 \ \rm V/nm$ in which both the exterior and the interior branches are in the $p$-wave phase.}
\label{cnt_10_SC}
\end{figure}
Denoting the length of the CNT by $L$, 
the boundary conditions, $\psi_{e}(x=0)=\psi_{e}(x=L)=0$, yield
the  restriction $\psi_{2K}^R(x) =-\psi_{2K'}^L(-x)$. Thus, the kinetic term
is given by $H_0^{(1)} = -i v_F \int_{-L}^{L}   \psi_{2K}^{R \dagger}(x)\partial_x    \psi_{2K}^{R }(x)$,
and the $p$-wave pairing term
between the exterior branches by
\begin{equation}
-\Delta_e \int_{-L}^{L} dx\, \text{sgn}(x)[ \psi_{2K}^R(x)\psi_{2K}^R(-x)+h.c.]\, .
\end{equation}
Solving for the zero energy mode  localized around $x=0$,
we obtain the MBS
\begin{equation}
\Psi_{e}^{M}(x)\propto \gamma_e\sin[(K+k_F)x ]e^{- x/\xi_e},\label{eq:Majorana}
\end{equation}
where $\gamma_e =\gamma_e^\dagger$,   and  it is assumed that the localization
length, given by $\xi_e =\hbar v_F/2\Delta_e$, satisfies $\xi_e \ll L $. 
Similarly for the interior branches, with the index $e$ replaced by $i$.

In general, the Majorana modes arising from the
interior and exterior branches at the same end of the CNT are not protected and can combine into a fermion.
To avoid such a scenario one needs to ensure the presence of  only one
single  MBS at each end of the CNT. This can be achieved in two ways.

First, there is a  window where the electric field $E$
can be chosen in such a way that the superconductivity in the
exterior  branch can be tuned into the $s$-wave phase, while the interior one still remains in the $p$-wave phase. In this case,
only the interior branch supports a MBS at each end of the nanotube, and we refer to this as a topological phase of the CNT (see dashed green line in Fig.~\ref{cnt_10_SC}a). 
Concretely, such a regime is  reached for $\Delta_e(k_F)> \delta > \Delta_i(k_F)$,
where $k_F\approx \sqrt{(\mu +eE\xi)^2 -\alpha^2/\hbar v_F}$ and  $\delta= eE\xi +\alpha -\mu$. With  Eq.~(\ref{super_coupling}) this criterion
becomes equivalent to $\Delta_{c2}\gtrsim\Delta_d \gtrsim \Delta_{c1}$, where $\Delta_{c1/c2}=\delta/(1\pm \sin \zeta)$.
For a given value of $\Delta_d$, the experimentally viable approach to drive the system into the topological phase is to tune the electric field $E$. Indeed,
 for $E_{c1}\gtrsim E \gtrsim E_{c2}$  (see Fig.~\ref{cnt_10_SC}b) 
 the exterior branch is in the  $s$-wave phase, while the interior one is in the $p$-wave phase.
 The critical value of the electric field $E_{c1}$ ($E_{c2}$) is determined by the condition $\delta = \Delta_e(k_F)$ ($\delta = \Delta_i(k_F)$). 
Similarly, we can tune the Fermi level.
[In passing we note that the gap $eE\xi$, and thus $\delta$, get  enhanced  by interaction effects around $k=0$
\cite{Braunecker_Jap_Klin_2009}, which is useful for experimental realizations. However, for simplicity we will ignore this feature here.]

Second, in the regime $\Delta_d  \lesssim \Delta_{c1}$ (see dotted blue line in Fig.~\ref{cnt_10_SC}a) or $E \gtrsim E_{c1}$  (see Fig.~\ref{cnt_10_SC}c)
 both branches are dominated by $p$-wave pairing. If the temperature is lower than both gaps, {\em i.e.} $k_B T<\Delta_{e,i}$, then there is an even number
 of MBS at each end of the nanotube, and the CNT is in the topologically trivial phase. However, in the intermediate regime with $\Delta_e> k_B T >\Delta_i$,
 the interior gap $\Delta_i$ is closed and the Majorana states  are removed, yet
those from the exterior branches remain, and the CNT is again in the topological phase. In the following we consider this latter scenario
 and discuss the role of interactions coming from the gapless states of the interior branch.

{\it Interactions.}
Interactions effects are most conveniently
described by linearizing  the spectrum of the fermionic fields $\psi_{2K}^{e/i}$ and $\psi_{2K'}^{e/i}$
near the Fermi momentum and expressing them in terms of the bosonized fields.
 The quadratic part of the  bosonized Hamiltonian thus obtained
has the  following form,
\begin{equation}
 \mathcal{H}_0= \frac{1}{2}\sum_{n=\pm}  \{\upsilon_nK_n (\partial_x \theta_n)^2+ \frac{\upsilon_n}{K_n} (\partial_x \phi_n)^2\},
\end{equation}
where  $\partial_x \phi_{+}$ and $\partial_x \phi_{-}$ are the sum and difference of
densities  between the two fermionic bands. The fields  conjugate to them are defined as,
$\theta_{+}$ and $\theta_{-}$, {\em resp.}  The  parameters $K_{+}\simeq 1- U_0/\pi \upsilon$ and $K_{-}\simeq 1+ (1-\left<S_z\right>^2) U_{2k_F}/2 \pi \upsilon$ encode information about the interactions and the renormalized velocities are given as $\upsilon_{+}\simeq\upsilon_F + U_0/\pi$, and $\upsilon_{-}\simeq\upsilon_F+b'(1+\left<S_z\right>^2)/4 \pi$, where the $b'$-term \cite{gogolin_egger_RG_CNT_1997,Kane_RG_CNT_1997} is due to the backscattering
contribution.  Here, $U_{0,2k_F}$ denotes the Fourier component of the screened Coulomb interaction.
Since $\left<S_z\right>^2<1$ and thus $K_->1$, we conclude \cite{giamarchi} that the forward scattering term $\propto \int dx d\tau \cos ({\sqrt{8\pi}\phi_{-}})$ 
scales to zero.

 Additional terms induced by  the proximity effect
lead to a modified Hamiltonian given by
\begin{equation}
 \mathcal{H} = \mathcal{H}_0 + \frac{\Delta_{e}}{2 \pi a } \cos \ \sqrt{2\pi} (\theta_+ - \phi_-)+ \frac{\Delta_{i}}{2 \pi a } \cos \ \sqrt{2\pi} (\theta_+ + \phi_-).
\end{equation}
Since we assume here $\Delta_e> k_B T >\Delta_i$, the second
term due to the interior branches (~$\Delta_i$) is smeared out by temperature effects
and will not be considered.

Using standard techniques \cite{giamarchi,tsvelik}, we derive the following renormalization group (RG) equations,
\begin{eqnarray}
\frac{dK_+}{dl}&=&\frac{f^2}{4}\left(1+ \frac{4\gamma  K_+ K_{-} }{(1+\gamma)^2 }  \right),\label{RG1}\\
\frac{dK_{-}^{-1}}{dl}&=&\frac{f^2}{4}  \left(1+     \frac{4\gamma }{K_{-}K_+(1+\gamma)^2 }  \right),\label{RG2}\\
\frac{d\gamma}{dl}&=&\frac{f^2}{4} \frac{\gamma (1-\gamma)K_{+}}{ (1+\gamma)(K_+K_{-}+1)}\, ,\label{RG3}\\
\frac{df}{dl}&=&f\left(2-\frac{1}{2K_+}-\frac{K_-}{2}\right) ,\label{RG4}
\end{eqnarray}
where  the flow parameter $l = \ln[a/a_0]$, $f=2\Delta_e a$, and $\gamma$ is the ratio of the velocities $\upsilon_+/\upsilon_-$.
We note that for the non-interacting case $\gamma$ is already at its fixed point,
$\gamma=1$, and including interactions (the repulsive interactions are
assumed to be well  screened)  causes only a small deviation from unity~\cite{starykh,klinovaja_unpub}.
Thus, it is convenient to assume $\gamma=1$, and under this assumption $K_{+} K_{-}$
is a constant, given in leading order by unity. Above RG equations now acquire the simple form $dR/dl= f^2/2$ and $df/dl=f(2-1/R)$,
where $R=(1/K_{+}+K_-)/2$. 
These  equations
are exactly the same  as in Ref.~\cite{suhas_majorana} derived  for interacting
spinless fermions in an effective $p$-wave regime. 
We conclude that for a  CNT
whose  initial values of the parameters lie in the  regime
$f_0 > 2 \sqrt{2 R_0 -\ln (2R_0 e)}$ has its RG flow such that both $K_{+}$ and $K_{-}$
approach the non-interacting value. At this point the problem can  be refermionized into a set of decoupled
gapped and gapless fermions and for a strongly screened CNT with initial value {\em e.g.} $K_+=0.8$ the localization length $\xi_e$ increases by $25\%$. 
Therefore, we conclude that the MBS which arise from  gapped fermions
 remain protected even in the presence of interacting gapless fermions and simply acquire a renormalized $\xi_e$.

{\it Conclusions.}
We have shown that an armchair CNT with helical
modes generated by an external electric field is a promising candidate material for
Majorana bound states.
By placing the CNT on top of an $s$-wave
superconductor  and  tuning the Fermi level and the electric field,  one can induce pairing
of Kramers partners from opposite Dirac points. This pairing  opens up  inequivalent gaps for the exterior
and the interior branches.  The Majorana modes obtained are stabilized  by either  
 tuning the electric field such that  the exterior gap acquires a predominantly
$s$-wave character or by increasing the temperature to remove the pairing in the interior
branches.

\emph{Acknowledgements.} We thank Karsten Flensberg for valuable discussions in the initial stage of
this work. We also acknowledge helpful discussions with Diego Rainis and
Pascal Simon. This work is supported by the Swiss NSF,
NCCR Nanoscience and NCCR QSIT, and DARPA.


\begin{thebibliography}{8}

\bibitem{fu} L. Fu and C. L. Kane, Phys. Rev. Lett. {\bf 100}, 096407 (2008).

\bibitem{Akhmerov} A. R. Akhmerov, J. Nilsson, and C. W. J. Beenakker
Phys. Rev. Lett. {\bf 102}, 216404 (2009).

\bibitem{lutchyn_majorana_wire_2010} R. M. Lutchyn, J. D. Sau, and S. Das Sarma,
Phys. Rev. Lett. {\bf 105}, 077001 (2010).

\bibitem{oreg_majorana_wire_2010} Y. Oreg, G. Refael, and F. von Oppen,
Phys. Rev. Lett. {\bf 105}, 177002 (2010).

\bibitem{potter_majoranas_2011} A. C. Potter and P. A. Lee,
Phys. Rev. B {\bf 83}, 094525 (2011).

\bibitem{alicea_majoranas_2010} J. Alicea, Phys. Rev. B {\bf 81}, 125318 (2010).

\bibitem{Qi}X. L. Qi and S. C. Zhang, Rev. Mod. Phys. {\bf 83}, 1057  (2011).

\bibitem{suhas_majorana}  S. Gangadharaiah, B. Braunecker, P. Simon, and
D. Loss, Phys. Rev. Lett. {\bf 107}, 036801 (2011).

\bibitem{stoudenmire}
E. M. Stoudenmire, J. Alicea, O. Starykh, and M.P.A. Fisher, Phys. Rev. B {\bf 84}, 014503 (2011).

\bibitem{lutchyn}
R. M. Lutchyn and M.P.A. Fisher, arXiv:1104.2358.


\bibitem{footnote_Overhauser_field} The Zeeman splitting can also be generated internally e.g. by the Overhauser field
coming from the hyperfine interaction between nuclear and electron spins \cite{Braunecker}.


\bibitem{stanescu}   T. D. Stanescu, R. M. Lutchyn, and S. Das Sarma,
Phys. Rev. B {\bf 84}, 144522 (2011).


\bibitem{klinovaja_helical_modes_2011} J. Klinovaja, M. J. Schmidt, B. Braunecker, and D. Loss,
Phys. Rev. Lett. {\bf 106}, 156809 (2011).

\bibitem{klinovaja_cnt_phys_rev_B} J. Klinovaja, M. J. Schmidt, B. Braunecker, and D. Loss,
Phys. Rev. B {\bf 84}, 085452 (2011).


\bibitem{footnote_k=0}
A recent proposal for MBS in CNTs focuses on the spectrum around the $\Gamma$-point \cite{Sau_CNT} (instead of the $K$, $K^\prime$-points with a Dirac spectrum considered here), where  this difficulty would be absent in principle. However, the effective Hamiltonian around the $\Gamma$-point  for a 
realistic CNT with SOI is not known and  needs separate treatment. 

\bibitem{le_hur_2008} K. Le Hur, S. Vishveshwara, and C. Bena,
Phys. Rev. B {\bf 77}, 041406 (2008).

\bibitem{izumida_soi_cnt_2009} W. Izumida, K. Sato, and R. Saito,
J. Phys. Soc. Jpn. {\bf 78}, 074707 (2009).

\bibitem{jeong_soi_cnt_2009} J.-S. Jeong and H.-W. Lee,
Phys. Rev. B {\bf 80}, 075409 (2009).

\bibitem{Dresselhaus} R. Saito, G. Dresselhaus, and M. S. Dresselhaus,
{\sl Physical Properties of Carbon Nanotubes} (Imperial College Press, London, 1998).

\bibitem{Braunecker_Jap_Klin_2009} B. Braunecker, G. I. Japaridze, J. Klinovaja, and D. Loss, 
Phys. Rev. B {\bf 82}, 045127 (2010).


\bibitem{gogolin_egger_RG_CNT_1997} R. Egger and A. Gogolin, Phys. Rev. Lett. {\bf 79}, 5082 (1997); 
Eur. Phys. J.  B {\bf 3}, 23 (1998).

\bibitem{Kane_RG_CNT_1997} C. Kane, L. Balents, and M.P.A. Fisher, Phys. Rev. Lett. {\bf 79}, 5086 (1997).


\bibitem{giamarchi} T. Giamarchi, {\sl Quantum Physics in One Dimension},
(Clarendon Press, Oxford, 2004).

\bibitem{tsvelik}
A. O. Gogolin, A. A. Nersesyan, and A. M. Tsvelik,
{\sl Bosonization and Strongly Correlated Systems},
(University Press, Cambridge, 1998).

\bibitem{starykh} O. A. Starykh, D. L. Maslov, W. Haeusler, and L. I.
Glazman, in Lecture Notes in Physics, ed. T. Brandes,
vol.~544, p.37, 1999.

\bibitem{klinovaja_unpub} J. Klinovaja, S. Gangadharaiah, and D. Loss, in preparation.


\bibitem{Braunecker} B. Braunecker, P. Simon, and D. Loss, Phys. Rev. B {\bf 80}, 165119 (2009).

\bibitem{Sau_CNT} J. D. Sau and S. Tewari, arXiv:1111.5622.

\end{thebibliography}
\end{document}